\documentclass[aps,reprint,superscriptaddress,article]{revtex4-2}
\usepackage{graphicx}
\usepackage{epstopdf}
\usepackage{bm}
\usepackage{amsmath}
\usepackage{amssymb}
\usepackage{xcolor}
\usepackage[colorlinks=true, letterpaper=true, pdfstartview=FitV, linkcolor=blue, citecolor=blue, urlcolor=blue]{hyperref}

\begin{document}

\title{Quantum Geometry Effects in Lattice Dynamics}

\author{Haoshu Li}
\email{lihaoshu@ustc.edu.cn}
\affiliation{Department of Physics, University of Science and Technology of China, Hefei 230026, China}

\author{Hao Chen}
\email{phych@njupt.edu.cn}
\affiliation{School of Science, Nanjing University of Posts and Telecommunications, Nanjing 210023, China}

\author{Qian Niu}
\affiliation{Department of Physics, University of Science and Technology of China, Hefei 230026, China}

\begin{abstract}
Quantum geometric effects such as the Berry curvature and the quantum metric are well established in electronic transport, whereas in lattice dynamics the Berry curvature also plays a key role: adiabatically following the lattice, electrons generate a molecular Berry curvature that splits degenerate optical phonons into chiral partners. To determine how the quantum metric affects phonons, we go beyond adiabatic lattice dynamics, identifying Berry curvature and quantum metric contributions to the electronic response of the lattice. Using first-principles calculations in the magnetic topological semimetal EuB$_6$, we find that phonon modes with chiral splitting and those with nonchiral splitting coexist. Unlike the former, which originate from the Berry curvature, the latter arise from the pure quantum metric, while the Berry curvature contribution is forbidden by crystal symmetry.
\end{abstract}

\maketitle

\textcolor{blue}{\textit{Introduction.}---}Recently, quantum geometry has become a foundational framework for understanding electronic transport. Transport depends on both the band dispersion and the variation of Bloch states across momentum space, and the quantum geometric tensor describes the local geometry of these states~\cite{GaoQuantumGeometryRMP2026,VermaQuantumGeometry2026}. The quantum geometric tensor consists of two parts: the real and symmetric part, known as the quantum metric, and the imaginary and antisymmetric part, known as the Berry curvature.
The Berry curvature determines the geometric phase accumulated around a closed path in momentum space. Under an applied electric field, it endows an electronic wave packet with a transverse anomalous velocity~\cite{SundaramNiuWavePacket1999}; integrating it over the Brillouin zone weighted by the electronic occupation yields the intrinsic anomalous Hall conductivity~\cite{XiaoChangNiu2010,NagaosaAnomalousHall2010,YaoAnomalousHall2004, GaoLayerHall2021}.
This mechanism is reflected in anomalous Hall measurements on the ferromagnetic kagome semimetal Co$_3$Sn$_2$S$_2$ and the noncollinear antiferromagnet Mn$_3$Sn~\cite{LiuKagomeHall2018,NakatsujiAnomalousHall2015}.
The quantum metric measures the distance between neighboring Bloch states~\cite{ProvostVallee1980}.
Electric field induced interband mixing brings an energy-weighted quantum metric into the second-order transport response. Where symmetry permits, this contribution generates an intrinsic nonlinear Hall current~\cite{PhysRevLett.112.166601,KaplanNonlinearGeometry2024,LiuIntrinsicSecondOrderHall2021}.
Nonlinear Hall responses associated with the quantum metric have been observed in topological antiferromagnets and their heterostructures~\cite{WangQuantumMetricTransport2023,GaoQuantumMetricHall2023}.

Berry curvature also affects lattice dynamics because the electronic ground-state wave function is parametrized by the atomic positions. In the Born--Oppenheimer approximation, the electrons follow the instantaneous lattice configuration~\cite{BornHuangCrystalLattices1996}.
When the ionic coordinates trace a closed path, the electronic state can acquire a geometric phase. The Berry curvature associated with this phase in ionic-displacement space is known as the molecular Berry curvature~\cite{MeadTruhlar1979,MeadMolecularGeometricPhase1992}. In the adiabatic lattice dynamics of a crystal, the molecular Berry curvature acts back on the lattice through a Lorentz-like geometric force proportional to the ionic velocity, modifying phonon frequencies and vibrational modes. When time-reversal symmetry is broken, this force can couple two degenerate linearly polarized optical modes to form chiral phonons with opposite directions of rotation and different frequencies~\cite{SaparovXiongRenNiu2022}.

Given that the Berry curvature of electronic states in ionic-displacement space can modify lattice dynamics, one may ask whether the quantum metric can produce a distinguishable effect as well. We address this question by going beyond the adiabatic approximation and identifying the Berry curvature and quantum metric contributions to the electronic response of the lattice. Using first-principles electron--phonon coupling matrix elements, we compare their effects on zone-center optical phonons in the ferromagnetic topological semimetal EuB$_6$. For the triply degenerate reference $T_{2u}$ mode, the imaginary, antisymmetric self-energy associated with the Berry curvature yields two chiral modes with oppositely directed phonon angular momenta. For the $E_g$ doublet, fourfold rotation about the magnetization axis forbids this coupling within the doublet, but allowing the real, symmetric self-energy associated with the quantum metric to split two rotation sectors. The split modes have nearly zero net phonon angular momentum within numerical accuracy. The different polarization patterns provide a way to distinguish pure metric contributions to lattice dynamics from Berry curvature contributions to lattice dynamics.

\textcolor{blue}{\textit{Lattice Dynamics Beyond the Adiabatic Approximation.}---}The quantum metric contribution requires a treatment beyond the adiabatic approximation. This is particularly relevant in metals and semimetals, where the electronic and vibrational time scales may be comparable. An electronic energy separation $E_{\mathrm{gap}}$ sets a characteristic evolution rate $E_{\mathrm{gap}}/\hbar$, as illustrated by the phase factor $e^{-i E_{\mathrm{gap}}t/\hbar}$, while the lattice evolves at the phonon frequency $\omega_{\mathrm{ph}}$. The adiabatic condition $E_{\mathrm{gap}}\gg\hbar\omega_{\mathrm{ph}}$ may fail for low-energy electronic transitions in metals and semimetals. We use ferromagnetic EuB$_6$ to illustrate the resulting frequency-dependent lattice response.

\begin{figure}[htbp]
  \includegraphics[width=\linewidth]{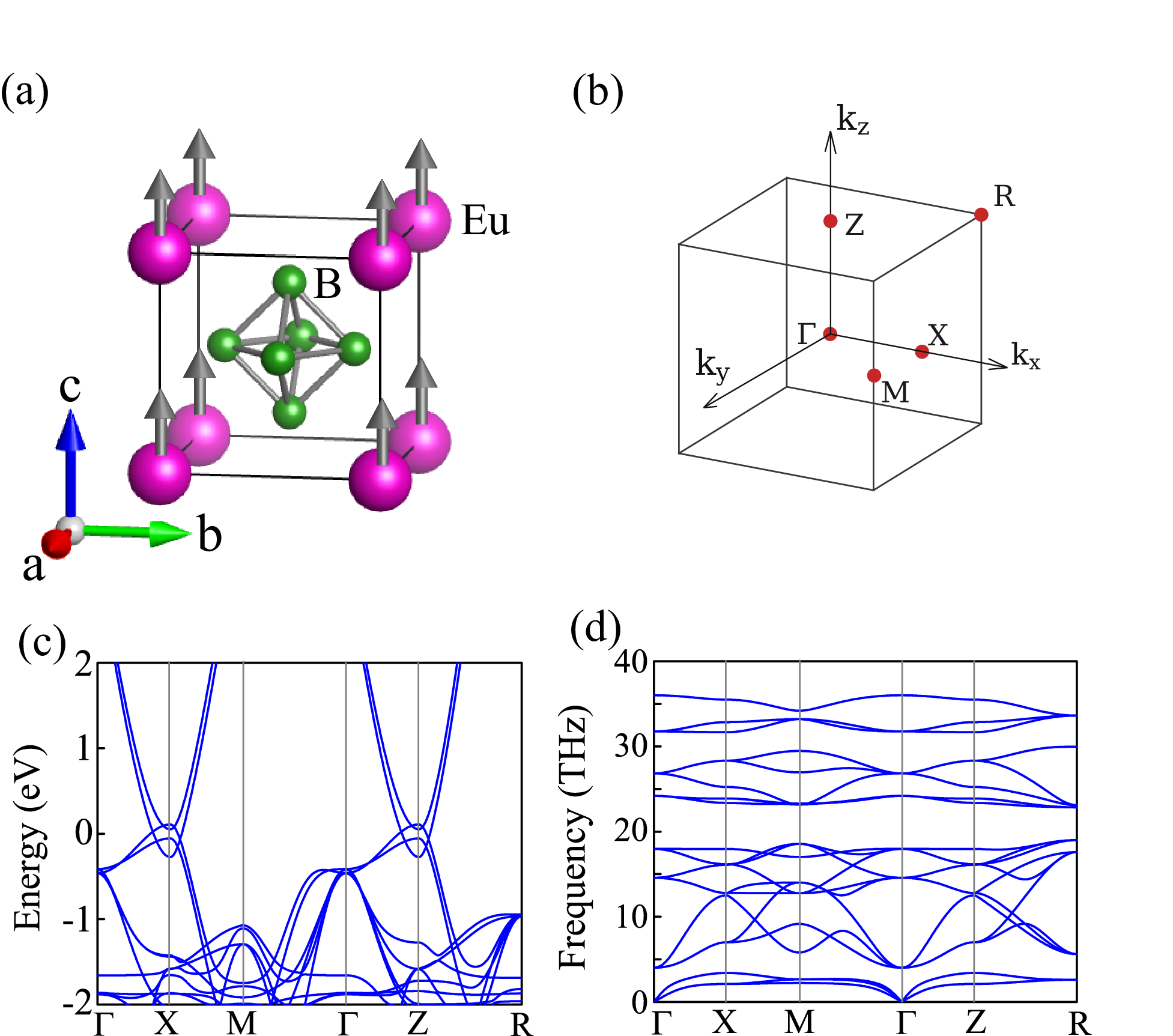}
  \caption{\label{fig1} Crystal structure, Brillouin zone, and electronic and phonon dispersions of ferromagnetic EuB$_6$.
  (a) Crystal structure. Purple and green spheres denote Eu and B atoms, respectively. Gray arrows indicate the Eu moments aligned along $[001]$.
  (b) First Brillouin zone and high-symmetry points $\Gamma$, X, M, Z, and R.
  (c) Electronic band structure along $\Gamma$--X--M--$\Gamma$--Z--R.
  (d) Phonon dispersion along the same path.}
\end{figure}

Figure~\ref{fig1}(a) shows the crystal structure of EuB$_6$. Ignoring magnetic order, the cubic crystal structure has point group $m\bar{3}m$. For the magnetic structure, the Eu moments are along $[001]$, which lowers the symmetry to the magnetic point group $4/mm'm'$, which retains fourfold rotation $C_{4z}$ about the magnetization axis. Figure~\ref{fig1}(b) shows the first Brillouin zone and the high-symmetry points. Figures~\ref{fig1}(c) and \ref{fig1}(d) show the electronic bands and reference phonon dispersion along the same path. The low-energy band crossings near X and Z are related to the previously reported nodal-ring electronic structure~\cite{PhysRevX.11.021016}.
The electronic transition energies in these regions need not be large compared with the phonon energies, motivating the nonadiabatic lattice dynamics. The triply and doubly degenerate reference modes at $\Gamma$ in Fig.~\ref{fig1}(d) provide two distinct symmetry subspaces in which to compare the nonadiabatic corrections.

The phonon frequencies are obtained self-consistently from the secular equation $\det\!\left[\omega^2 - D(\bm q) - 2\tilde{\Pi}(\bm q,\omega)\right]=0$, where $D(\bm q)$ is the dynamical matrix and $\tilde{\Pi}(\bm q,\omega)$ is the frequency-dependent phonon self-energy. We use $\tilde{\Pi}$ to distinguish this quantity from the conventional phonon self-energy $\Pi$ because their definitions differ by a factor with units of frequency \cite{RevModPhys.89.015003, supplemental}.
\begin{align}
    \tilde{\Pi}_{\kappa \alpha, \kappa' \beta}(\bm q, \omega) 
    = & \frac{1}{2\sqrt{M_{\kappa} M_{\kappa'}}}\sum_{mn}\int_{\text{BZ}} \frac{d \bm{k}}{\Omega_{\text{BZ}}} g_{mn, \kappa \alpha} g^*_{mn, \kappa' \beta} \notag \\
    &\quad\times \left( \frac{f_{mn}}{\varepsilon_{mn} - \hbar \omega} - \frac{f_{mn}}{\varepsilon_{mn}} \right),
    \label{eq: self_energy}
\end{align}
Here $g_{mn,\kappa\alpha}\equiv g_{mn,\kappa\alpha}(\bm k,\bm q)$ is the electron--phonon coupling matrix element, i.e.,
\begin{align}
g_{mn,\kappa\alpha}
  &= \sum_p e^{i\bm q\cdot\bm R_p} \times
  \left\langle u_{m\bm{k}+\bm{q}} \left|
  \frac{\partial V(\bm{r};\{\bm{\tau}_{\kappa'' p''}\})}
       {\partial \bm{\tau}_{\kappa p,\alpha}}
  \right| u_{n\bm{k}}\right\rangle .
\end{align}
The occupation and energy differences are $f_{mn}\equiv f_{m\bm{k}+\bm{q}}-f_{n\bm{k}}$ and $\varepsilon_{mn}\equiv\epsilon_{m\bm{k}+\bm{q}}-\epsilon_{n\bm{k}}$, respectively. The phonon eigenvector $\psi$ is a null vector of the matrix in the secular equation, $\left[\omega^2-D(\bm q)-2\tilde{\Pi}(\bm q,\omega)\right]\psi=0$. The dressed-phonon derivation gives the following normalization condition \cite{supplemental}.
\begin{align}
	\psi_{\nu}^\dag(\bm q) \left[1 - \frac{1}{\omega}\frac{\partial \tilde{\Pi}(\omega)}{\partial \omega}\Big|_{\omega_{\nu, \bm q}} \right] \psi_{\nu}(\bm q) = 1.
    \label{eq: wavef_normalization}
\end{align}
With this normalization, the expression for the phonon angular momentum along $z$ retains the form $L_{\text{ph}, \nu\bm{q}}^z=\psi_\nu^\dag(\bm q)\hat{S}^z\psi_\nu(\bm q)$, where $\hat{S}^z$ follows Ref.~\cite{Zhang2015-PRL}.
The self-energy thus determines the spectral corrections, and the corresponding normalized eigenvectors determine the phonon angular momenta.

\textcolor{blue}{\textit{Quantum Geometry in Lattice Dynamics.}---}To identify the contributions beyond the adiabatic molecular Berry curvature, we decompose the phonon self-energy into parts even and odd in frequency,
$\tilde{\Pi}_{\kappa\alpha,\kappa'\beta}=\tilde{\Pi}_{\kappa\alpha,\kappa'\beta}^{\mathrm{even}}+\tilde{\Pi}_{\kappa\alpha,\kappa'\beta}^{\mathrm{odd}}$, with
\begin{align}
	\tilde{\Pi}_{\kappa \alpha, \kappa' \beta}^{\text{even}}(\bm q, \omega) 
    = & \frac{1}{2\sqrt{M_{\kappa} M_{\kappa'}}}\sum_{mn}\int_{\text{BZ}} \frac{d \bm{k}}{\Omega_{\text{BZ}}} g_{mn, \kappa \alpha} g^*_{mn, \kappa' \beta} \notag \\
    &\quad\times \left[\frac{f_{mn} \varepsilon_{mn}}{\varepsilon_{mn}^2 - (\hbar \omega)^2} - \frac{f_{mn}}{\varepsilon_{mn}}\right],  \notag \\
	\tilde{\Pi}_{\kappa \alpha, \kappa' \beta}^{\text{odd}}(\bm q, \omega) 
    = & \frac{1}{2\sqrt{M_{\kappa} M_{\kappa'}}}\sum_{mn}\int_{\text{BZ}} \frac{d \bm{k}}{\Omega_{\text{BZ}}} g_{mn, \kappa \alpha} g^*_{mn, \kappa' \beta} \notag \\
    &\quad\times \left[\frac{f_{mn} (\hbar \omega)}{\varepsilon_{mn}^2 - (\hbar \omega)^2} \right].
	\label{eq: decom}
\end{align}

The leading term in the low-frequency expansion of $\tilde{\Pi}^{\mathrm{odd}}$ is linear in frequency and has the form
\begin{align}
&\frac{\hbar \omega}{2\sqrt{M_{\kappa} M_{\kappa'}}}
 \sum_{\substack{\varepsilon_{m}<\mu,\\ \varepsilon_{n}>\mu}}\int_{\text{BZ}} \frac{d\bm k}{\Omega_{\text{BZ}}} \notag \\
&\qquad\times
 \left[
 \frac{g_{mn,\kappa\alpha}g^*_{mn,\kappa'\beta}}{\epsilon_{mn}^2}
 -
 \frac{g^*_{nm,\kappa'\beta}g_{nm,\kappa\alpha}}{\epsilon_{nm}^2}
 \right].
\end{align}
This linear term recovers the adiabatic molecular Berry curvature. 

The leading frequency-even term is quadratic in $\omega$ and has the form
\begin{align}
&\frac{(\hbar\omega)^2}{2\sqrt{M_{\kappa} M_{\kappa'}}}
 \sum_{mn}\int_{\text{BZ}} \frac{d\bm k}{\Omega_{\text{BZ}}} 
\,
 2g_{mn,\kappa\alpha}g^*_{mn,\kappa'\beta}
 \frac{f_{mn}}{\epsilon_{mn}^3}.
\end{align}
Replacing the electron--phonon coupling matrix elements $g_{mn,\kappa\alpha}$ by matrix elements of the electronic velocity operator $(v_{\alpha})_{mn}$ gives an expression analogous to the nonadiabatic metric in electronic transport~\cite{pwvz-868b, PhysRevLett.112.166601}.
In a two-band model, this quantity is proportional to the quantum metric density~\cite{PhysRevLett.112.166601}.
Therefore, we refer to the odd and even parts below as the Berry curvature and quantum metric contributions, respectively. In a real basis at $\Gamma$, the frequency-odd part is purely imaginary and antisymmetric, whereas the frequency-even part is real and symmetric.

\begin{figure}[htbp]
  \includegraphics[width=\linewidth]{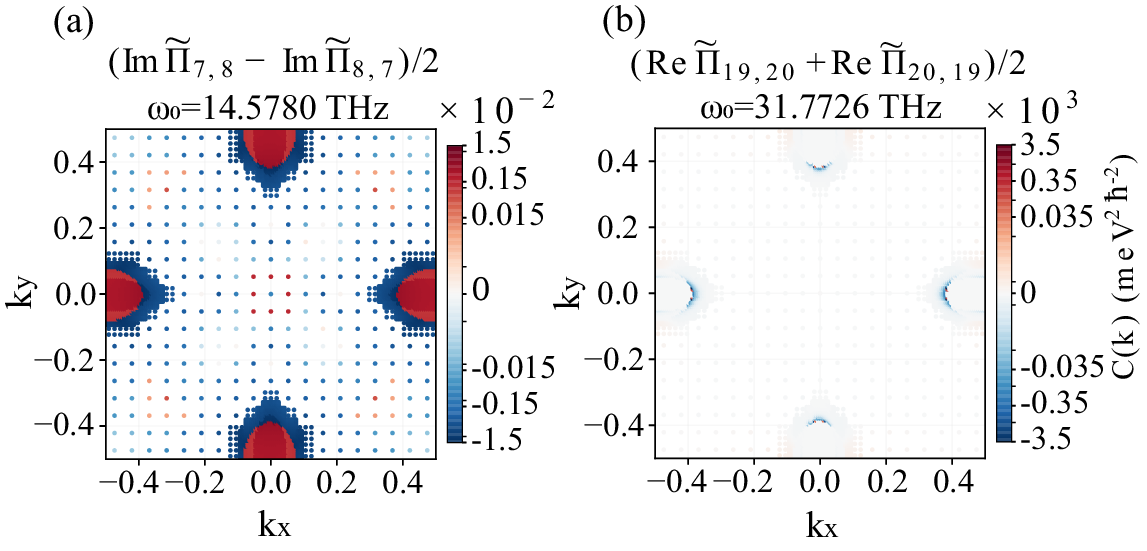}
  \caption{\label{fig2} Momentum-resolved contributions to the $\Gamma$-point phonon self-energy of EuB$_6$ in the $k_z=0$ plane.
  (a) Imaginary antisymmetric component between modes 7 and 8,
  $[\mathrm{Im}\widetilde{\Pi}_{7,8}-\mathrm{Im}\widetilde{\Pi}_{8,7}]/2$,
  evaluated at $14.5780~\mathrm{THz}$.
  (b) Real symmetric component between modes 19 and 20,
  $[\mathrm{Re}\widetilde{\Pi}_{19,20}+\mathrm{Re}\widetilde{\Pi}_{20,19}]/2$,
  evaluated at $31.7726~\mathrm{THz}$.
  Both panels use $E_F=4.2163~\mathrm{eV}$ and the reference $\Gamma$-point phonon-mode basis.
  The points are sampled nonuniformly in electronic momentum. Color denotes the local contribution $C(\mathbf{k})$ in $\mathrm{meV}^2 \hbar^{-2}$ before Brillouin-zone weights are applied. Red and blue indicate positive and negative contributions, respectively.
  The panels use independently scaled logarithmic color maps for positive and negative values. The coordinates $k_x$ and $k_y$ are given in reciprocal-lattice units.}
\end{figure}

Figure~\ref{fig2} illustrates the decomposition for two sets of $\Gamma$-point phonons. In the reference phonon-mode basis, Fig.~\ref{fig2}(a) shows the imaginary antisymmetric component between modes 7 and 8, and Fig.~\ref{fig2}(b) shows the real symmetric component between modes 19 and 20. These components probe the frequency-odd and frequency-even responses, respectively. We evaluate them at the corresponding reference frequencies of $14.5780$ and $31.7726~\mathrm{THz}$, with the Fermi level fixed at $E_F=4.2163~\mathrm{eV}$. We obtain these momentum-resolved self-energies from Wannier-interpolated, first-principles electron--phonon coupling matrix elements. The two panels show individual matrix elements in the reference mode basis. Both panels show pronounced variations in ring-shaped regions near X and Y, close to the low-energy band crossings discussed in connection with Fig.~\ref{fig1}(c). The color scale preserves the sign of the contribution at each momentum. The integral of the momentum-resolved self-energy is the phonon self-energy in Eq.~(\ref{eq: self_energy}) and is evaluated numerically as a discrete sum over grid points, which counts the nonuniform weights of the points on the nonuniform mesh.

\begin{figure*}[htbp]
  \includegraphics[width=\linewidth]{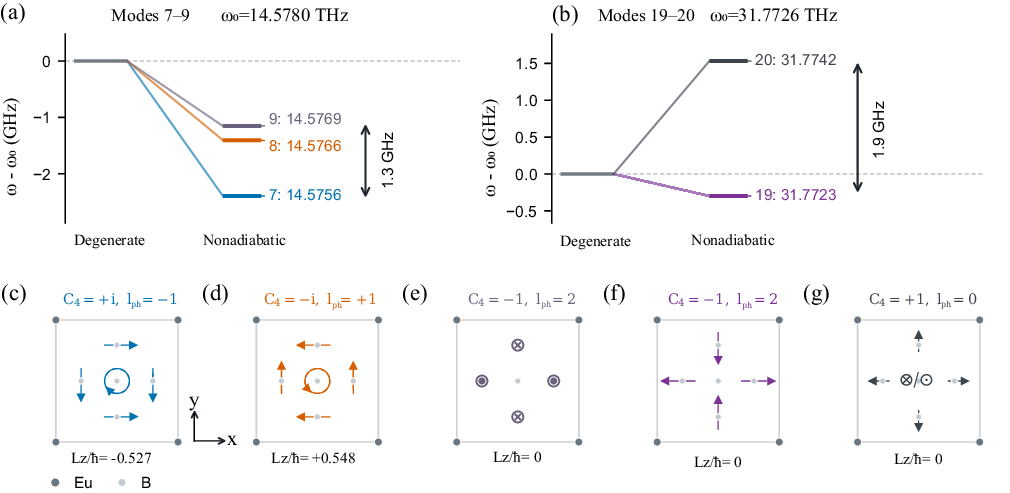}
  \caption{\label{fig3} Splitting and vibrational patterns of $\Gamma$-point phonons in EuB$_6$ due to nonadiabatic lattice dynamics.
  (a) Split frequencies of modes 7--9 relative to a common reference frequency of $14.5780~\mathrm{THz}$. The maximum frequency separation is about $1.3~\mathrm{GHz}$.
  (b) Split frequencies of modes 19 and 20 relative to a common reference frequency of $31.7726~\mathrm{THz}$. The two modes are separated by about $1.9~\mathrm{GHz}$.
  Dashed lines mark the reference frequencies, and connecting lines guide the eye.
  (c)--(e) Vibrational patterns of modes 7, 8, and 9, respectively.
  (f),(g) Vibrational patterns of modes 19 and 20, respectively.
  Patterns are projected onto the $xy$ plane. Dark and light gray dots mark the equilibrium positions of Eu and B, respectively.
  Curves trace the atomic paths, arrows indicate the direction of motion at the chosen initial phase, and $\odot$ and $\otimes$ denote motion along $+z$ and $-z$, respectively.
  Each panel gives the fourfold rotation eigenvalue $\lambda_{C_4}$ and phonon pseudoangular momentum $l_{\mathrm{ph}}$, defined by $\lambda_{C_4}=\exp(-i\pi l_{\mathrm{ph}}/2)$, along with the phonon angular momentum along $z$, $L_z/\hbar$.
  The Fermi level is fixed at $E_F=4.2163~\mathrm{eV}$.}
\end{figure*}

We solve the secular equation with the frequency-dependent self-energy and normalize the phonon eigenvectors using Eq.~(\ref{eq: wavef_normalization}). Figure~\ref{fig3} shows the resulting frequencies and vibrational patterns of modes 7--9 and 19--20. In Fig.~\ref{fig2}, the self-energy is evaluated at fixed reference frequencies. The frequencies in Fig.~\ref{fig3} instead follow from the self-consistent solution. The lines connecting the frequencies to each group's common reference are guides to the eye.

\textcolor{blue}{\textit{Phonon Splitting Induced by Berry Curvature.}---}For $T_{2u}$ modes 7--9 in Fig.~\ref{fig3}(a), the three nonadiabatic frequencies are $14.5756$, $14.5766$, and $14.5769~\mathrm{THz}$, all below the common reference frequency of $14.5780~\mathrm{THz}$. At the precision shown in the figure, the separation across the three branches is about $1.3~\mathrm{GHz}$, and the splitting between the two chiral branches, modes 7 and 8, is about $1.0~\mathrm{GHz}$.

Figures~\ref{fig3}(c) and \ref{fig3}(d) show the atomic trajectories for modes 7 and 8. The modes rotate in opposite directions, with angular momenta along the magnetization axis of $L_z/\hbar=-0.527$ and $+0.548$, respectively. Their fourfold rotation eigenvalues are $+i$ and $-i$, corresponding to pseudoangular momenta $l_{\mathrm{ph}}=-1$ and $+1$ under the convention $\lambda_{C_4}=\exp(-i\pi l_{\mathrm{ph}}/2)$. Mode 9 in Fig.~\ref{fig3}(e) has a real rotation eigenvalue of $-1$, pseudoangular momentum $l_{\mathrm{ph}}=2$, and $L_z=0$. The pseudoangular momentum specifies the phase acquired under a crystal rotation, while the phonon angular momentum describes the motion of the ions.

Within this threefold-degenerate phonon wavefunction subspace, both the antisymmetric Berry curvature and the symmetric quantum metric components of the phonon self-energy contribute to the nonadiabatic lattice dynamics; that is, this constitutes a mixed effect. Specifically, the imaginary antisymmetric self-energy mixes the two degenerate vibration components into complex combinations carrying opposite angular momenta. The real symmetric self-energy further shifts the mode frequencies and can give the nonchiral mode and the chiral pair different overall shifts. The unequal spacings in Fig.~\ref{fig3}(a) therefore reflect both contributions. If the quantum metric contribution vanished, the three bands would be equally spaced, with the nonchiral mode lying midway between the two opposite chiral modes. In reality, however, the three bands are not equally spaced, and the nonchiral mode even lies outside the frequency interval spanned by the two opposite chiral modes. This indicates that the quantum metric contribution is of the same order of magnitude as the Berry curvature contribution, and that its effect on the frequencies is even larger than that of the Berry curvature. Nevertheless, it is the Berry curvature that endows the modes with chirality. Figures~\ref{fig3}(c)--\ref{fig3}(e) show the coordinated atomic motion and relative phases of the three eigenmodes.

\textcolor{blue}{\textit{Phonon Splitting Induced by the Quantum Metric.}---}The doubly degenerate $E_g$ modes 19 and 20 also split [Fig.~\ref{fig3}(b)], and both have $L_z=0$ within numerical accuracy [Figs.~\ref{fig3}(f) and \ref{fig3}(g)]. Their nonadiabatic frequencies are $31.7723$ and $31.7742~\mathrm{THz}$, respectively, slightly below and above the common reference of $31.7726~\mathrm{THz}$. The two branches are separated by about $1.9~\mathrm{GHz}$. Thus a finite splitting can occur with vanishing phonon angular momentum in both modes.

The $C_{4z}$ symmetry accounts for this result. In the $E_g$ subspace spanned by $2z^2-x^2-y^2$ and $x^2-y^2$, fourfold rotation is represented by $\mathrm{diag}(+1,-1)$. The Brillouin zone integrated self-energy commutes with this matrix, so its off-diagonal entries vanish in the symmetry-adapted basis. The imaginary antisymmetric coupling within the doublet therefore vanishes, while the real symmetric self-energy can have unequal diagonal entries, producing a nonchiral splitting. Of course, whether the real symmetric self-energy can indeed have unequal diagonal entries should also be confirmed to be allowed by elements of the magnetic point group other than $C_{4z}$; this is verified in the Supplementary Materials \cite{supplemental}. The real symmetric off-diagonal contribution in Fig.~\ref{fig2}(b) is expressed in the reference mode basis. 

Mode 19 in Fig.~\ref{fig3}(f) has $\lambda_{C_4}=-1$ and $l_{\mathrm{ph}}=2$, and mode 20 in Fig.~\ref{fig3}(g) has $\lambda_{C_4}=+1$ and $l_{\mathrm{ph}}=0$. Both eigenvectors can be chosen real and describe standing vibrations with no net phonon angular momentum. Under fourfold rotation, $l_{\mathrm{ph}}=2$ and $-2$ are equivalent modulo four, so the pseudoangular momentum of mode 19 specifies no preferred direction of rotation.

Together, $T_{2u}$ modes 7--9 and $E_g$ modes 19--20 show how Berry curvature related antisymmetric mixing produces chiral modes and how a pure quantum metric related symmetric mixing permits nonchiral splitting.

\textcolor{blue}{\textit{Discussion and Conclusion.}---}It is worth  noting that, unlike the Berry curvature contribution, which is time-reversal odd, the quantum metric can still yield a nonzero contribution in time-reversal preserving nonmagnetic systems. However, in time-reversal preserving systems, since the crystal symmetry is not broken before and after considering the quantum metric effect of electron–phonon coupling, the phonon level degeneracy imposed by the crystal symmetry remains intact, and the phonon spectrum exhibits only a trivial shift without any change in shape. Therefore, the nonchiral splitting discussed here can be observed only in time-reversal broken systems.

Of the 32 crystallographic point groups (ignoring magnetic order first), and further restricting to ferromagnetic configurations, only three cubic groups $432$, $\bar{4}3m$, and $m\bar{3}m$ meet the symmetry conditions for both types of splitting (chiral and nonchiral splittings) to occur in the same system~\cite{supplemental}. The splitting of the threefold-degenerate phonons discussed earlier is a chiral splitting, and its magnitude receives contributions from both the Berry curvature and the quantum metric. For such lattice dynamics with joint contributions, another useful fact is that, among all systems permitting a nonzero molecular Berry curvature, only cubic crystal systems can simultaneously possess a quantum metric capable of modifying the splitting magnitude. We consider this symmetry based observation, together with the adiabatic condition set by the comparison between the electronic gap and the phonon frequency, to be convenient for determining when the influence of the quantum metric must be taken into account together with the molecular Berry curvature.


By retaining the frequency dependence of the phonon self-energy, we identify two manifestations of electronic quantum geometry in lattice dynamics. In a real $\Gamma$-point mode basis, mixes degenerate vibrations into modes of opposite chirality, whereas the real symmetric contribution lifts the degeneracy of modes carrying zero net phonon angular momentum. Our first-principles calculations for ferromagnetic EuB$_6$ illustrate both types of splitting in the same material. The chiral optical modes 7 and 8 have opposite phonon angular momenta along the magnetization axis and are separated by about $1.0~\mathrm{GHz}$. The initially doubly degenerate $E_g$ modes 19 and 20 split by about $1.9~\mathrm{GHz}$, with zero phonon angular momentum for both. This identifies a route by which the quantum metric lifts a phonon degeneracy through a nonadiabatic frequency correction while preserving nonchiral vibrations. These results extend the role of the quantum metric from transport to nonadiabatic lattice dynamics and show how phonon frequencies and vibrational polarizations can distinguish the two geometric contributions.

\textcolor{blue}{\textit{Acknowledgement.}---}This work was supported by the National Natural Science Foundation of China (Grants No.~12504013 and No.~12234017) and the Basic Research Program of Jiangsu (Grant No.~BK20250647).


\bibliography{ChiraPhase}

\end{document}